\documentclass[12pt]{article}
\usepackage{graphicx}
\usepackage{lscape}
\usepackage{citesort}

\begin{document}

\def\ds{\displaystyle}
\def\beq{\begin{equation}}
\def\eeq{\end{equation}}
\def\bea{\begin{eqnarray}}
\def\eea{\end{eqnarray}}
\def\beeq{\begin{eqnarray}}
\def\eeeq{\end{eqnarray}}
\def\ve{\vert}
\def\vel{\left|}
\def\ver{\right|}
\def\nnb{\nonumber}
\def\ga{\left(}
\def\dr{\right)}
\def\aga{\left\{}
\def\adr{\right\}}
\def\lla{\left<}
\def\rra{\right>}
\def\rar{\rightarrow}
\def\nnb{\nonumber}
\def\la{\langle}
\def\ra{\rangle}
\def\ba{\begin{array}}
\def\ea{\end{array}}
\def\tr{\mbox{Tr}}
\def\ssp{{\Sigma^{*+}}}
\def\sso{{\Sigma^{*0}}}
\def\ssm{{\Sigma^{*-}}}
\def\xis0{{\Xi^{*0}}}
\def\xism{{\Xi^{*-}}}
\def\qs{\la \bar s s \ra}
\def\qu{\la \bar u u \ra}
\def\qd{\la \bar d d \ra}
\def\qq{\la \bar q q \ra}
\def\gGgG{\la g^2 G^2 \ra}
\def\q{\gamma_5 \not\!q}
\def\x{\gamma_5 \not\!x}
\def\g5{\gamma_5}
\def\sb{S_Q^{cf}}
\def\sd{S_d^{be}}
\def\su{S_u^{ad}}
\def\ss{S_s^{??}}
\def\sbp{{S}_Q^{'cf}}
\def\sdp{{S}_d^{'be}}
\def\sup{{S}_u^{'ad}}
\def\ssp{{S}_s^{'??}}
\def\sig{\sigma_{\mu \nu} \gamma_5 p^\mu q^\nu}
\def\fo{f_0(\frac{s_0}{M^2})}
\def\ffi{f_1(\frac{s_0}{M^2})}
\def\fii{f_2(\frac{s_0}{M^2})}
\def\O{{\cal O}}
\def\sl{{\Sigma^0 \Lambda}}
\def\es{\!\!\! &=& \!\!\!}
\def\ap{\!\!\! &\approx& \!\!\!}
\def\ar{&+& \!\!\!}
\def\ek{&-& \!\!\!}
\def\kek{\!\!\!&-& \!\!\!}
\def\cp{&\times& \!\!\!}
\def\se{\!\!\! &\simeq& \!\!\!}
\def\eqv{&\equiv& \!\!\!}
\def\kpm{&\pm& \!\!\!}
\def\kmp{&\mp& \!\!\!}


\def\simlt{\stackrel{<}{{}_\sim}}
\def\simgt{\stackrel{>}{{}_\sim}}


\title{
         {\Large
                 {\bf
$\Lambda_b \rar \Lambda \ell^+ \ell^-$ decay in universal extra dimensions
                 }
         }
      }

\author{\vspace{1cm}\\
{\small T. M. Aliev \thanks
{e-mail: taliev@metu.edu.tr}~\footnote{permanent address:Institute
of Physics,Baku,Azerbaijan}\,\,,
M. Savc{\i} \thanks
{e-mail: savci@metu.edu.tr}} \\
{\small Physics Department, Middle East Technical University,
06531 Ankara, Turkey} }

\date{}

\begin{titlepage}
\maketitle
\thispagestyle{empty}

\begin{abstract}
We study the exclusive $\Lambda_b \rar \Lambda \ell^+ \ell^-$ decay 
in the Appelquist, Chang, Dobrescu model within one universal extra
dimension. We investigate the sensitivity of the branching ratio, lepton
polarization and forward--backward asymmetry ${\cal A}_{FB}$ to the
compactification parameter $1/R$ We obtain that the branching ratio changed
about $50\%$ compared to the SM value, when $1/R=200~GeV$ and zero position
of the forward--backward asymmetry is shifted to the left compared to the SM
result. Therefore measurement of the branching ratio of $\Lambda_b \rar
\Lambda \ell^+ \ell^-$ decay and determination of zero position of ${\cal
A}_{FB}$ are very useful in looking for new physics in the framework of the 
UED models.
\end{abstract}

~~~PACS numbers: 12.60.--i, 13.30.--a, 14.20.Mr
\end{titlepage}

\section{Introduction}
Flavor--changing neutral current (FCNC) $b \rar s(d) \ell^+ \ell^-$
transitions are forbidden in the standard model (SM) at tree level that occur 
at loop level, and therefore provide consistency check of the SM at quantum
level. These decays induced by the FCNC are also very sensitive to the new
physics beyond the SM. New physics embedded into rare decays through the
Wilson coefficients which can take values different from their SM
counterpart or through the new operator structures in an effective
Hamiltonian (see \cite{R7601} and references therein).

Among the hadronic, leptonic and semileptonic decays, the last decay
channels are very significant, since they are theoretically, more or less,
clean, and they have relatively larger branching ratio. From theoretical
side there are many works in which the semileptonic decay channels due to
$b \rar s(d) \ell^+ \ell^-$ transitions are investigated. These decays contain 
many observables like forward--backward asymmetry ${\cal A}_{FB}$, lepton
polarization asymmetries, etc., which are very useful and serve as a testing
ground for the SM and looking for new physics beyond the SM \cite{R7601}.
From experimental side, BELLE \cite{R7602,R7603} and BaBar
\cite{R7604,R7605} collaborations provide recent measurements of the
branching ratios of the semileptonic decays due to the $b \rar s \ell^+
\ell^-$ transitions, which can be summarized as:
\bea
{\cal B}(B \rar K^\ast \ell^+ \ell^-) \es \left\{ \begin{array}{lc}
\left( 16.5^{+2.3}_{-2.2} \pm 0.9 \pm 0.4\right) \times
10^{-7}& \cite{R7602}~,\\ \\
\left( 7.8^{+1.9}_{-1.7} \pm 1.2\right) \times
10^{-7}& \cite{R7604}~,\end{array} \right.
\nnb \\ \nnb \\
{\cal B}(B \rar K \ell^+ \ell^-) \es \left\{ \begin{array}{lc}
\left( 5.5^{+0.75}_{-0.70} \pm 0.27 \pm 0.02\right) \times
10^{-7}& \cite{R7602}~,\\ \\
\left( 3.4  \pm 0.7 \pm 0.3\right) \times
10^{-7}& \cite{R7604}~.\end{array} \right.
\nnb \\ \nnb \\
{\cal B}(B \rar X_s \ell^+ \ell^-) \es \left\{ \begin{array}{lc}
\left( 4.11 \pm 0.83^{+0.85}_{-0.81}\right) \times
10^{-6}& \cite{R7603}~,\\ \\
\left( 5.6  \pm 1.5 \pm 0.6 \pm 1.1 \right) \times
10^{-6}& \cite{R7605}~.\end{array} \right. \nnb
\eea

Another exclusive decay which is described at inclusive level by the $b \rar
s \ell^+ \ell^-$ transition is the baryonic $\Lambda_b \rar \Lambda \ell^+
\ell^-$ decay. Unlike mesonic decays, the baryonic decays could maintain the
helicity structure of the effective Hamiltonian for the $b \rar s$
transition \cite{R7606}. Radiative and semileptonic decays of $\Lambda_b$
such as $\Lambda_b \rar \Lambda \gamma$, $\Lambda_b \rar \Lambda_c \ell
\bar{\nu}_\ell$, $\Lambda_b \rar \Lambda \ell^+ \ell^-$ $(\ell = e,\mu,\tau)$
and $\Lambda_b \rar \Lambda \nu \bar{\nu}$ have been extensively studied in
the literature \cite{R7607}--\cite{R7616}. More about heavy baryons, 
including the experimental prospects, can be found in \cite{R7607,R7618}. 

Among the various models of physics beyond the SM, extra dimensions
attract special interest, because they include gravity in addition other
interactions, giving hints on the hierarchy problem and a connection with
string theory. The model of Appelquist, Cheng and Dobrescu (ACD)
\cite{R7619} with one universal extra dimension (UED), where all the SM
particles can propagate in the extra dimension, are very attractive.
Compactification of the extra dimension leads to Kaluza--Klein (KK) model
in the four--dimension. In this model the only additional free parameter
with respect to the SM is $1/R$, i.e., inverse of the compactification
radius.

The restrictions imposed on UED are examined in the current accelerators,
for example, Tevatron experiments put the bound about $1/R\ge 300~GeV$.
Analysis of the anomalous magnetic moment \cite{R7620}, and $Z \rar \bar{b}
b$ vertex \cite{R7621} also lead to the bound $1/R\ge 300~GeV$.

Possible manifestation of UED models in the $K_L$--$K_S$ mass difference,
parameter $\varepsilon_K$, $B$--$\bar{B}_0$ mixing, $\Delta M_{d,s}$ mass
difference, and rare decays $K^+ \rar \pi \bar{\nu} \nu$ , $K_L \rar \pi^0
\bar{\nu} \nu$, $K_L \rar \mu^+ \mu^-$, $B \rar X_{s,d} \bar{\nu} \nu$, 
$B_{s,d} \rar \mu^+ \mu^-$, $B \rar X_s \gamma$, $B \rar X_s ~gluon$,
$B \rar X_s \mu^+ \mu^-$ and $\varepsilon^\prime/\varepsilon$ are
comprehensively investigated in \cite{R7622} and \cite{R7623}. Exclusive 
$B \rar K^\ast \ell^+ \ell^-$, $B \rar K^\ast \bar{\nu} \nu$ and $B \rar
K^\ast \gamma$ decays are studied in the framework of the UED scenario in
\cite{R7624}.

In the present work we study the $\Lambda_b \rar \Lambda \ell^+ \ell^-$
decay in the UED model. The plan of the paper is as follows. In section
2 we briefly discuss the main ingredients of ACD model and
study the rare $\Lambda_b \rar \Lambda \ell^+ \ell^-$ decay in it.
Section 3 is devoted to the numerical analysis and conclusions.

\section{Theoretical background for the $\Lambda_b \rar \Lambda \ell^+
\ell^-$ decay in universal extra dimension model}

Before presenting a detailed derivation of the matrix element of $\Lambda_b
\rar \Lambda \ell^+ \ell^-$ decay, let us discuss the main ingredients of
ACD model, which is the minimal extension of the SM in $4+\delta$
dimensions, and we consider the simplest case $\delta=1$.    
The five--dimensional ACD model with a single UED uses orbifold
compactification. The fifth dimension that is compactified in a circle of
radius $R$, with points $y=0$ and $y=\pi R$ that are fixed points of the
orbifolds. Generalization to the SM is realized by the propagating fermions, 
gauge bosons and the Higgs fields in all five dimensions. The Lagrangian can
be written as
\bea
{\cal L} = \int d^4x dy \Big\{ {\cal L}_A + {\cal L}_H + {\cal L}_F +
{\cal L}_Y \Big\}~, \nnb
\eea
where
\bea
{\cal L}_A \es - \frac{1}{4} W^{MNa} W_{MN}^a - 
\frac{1}{4} B^{MN} B_{MN}~, \nnb \\   
{\cal L}_H \es \Big( {\cal D}^M \phi \Big)^\dagger {\cal D}_M \phi 
- V(\phi)~, \nnb \\
{\cal L}_F \es \bar{ {\cal Q} } \Big( i \Gamma^M {\cal D}_M \Big) {\cal Q} +
\bar{u} \Big( i \Gamma^M {\cal D}_M \Big) u +
\bar{ {\cal D} } \Big( i \Gamma^M {\cal D}_M \Big) {\cal D}~, \nnb \\
{\cal L}_Y \es - \bar{ {\cal Q} } \widetilde{Y}_u \phi^c u - 
\bar{ {\cal Q} } \widetilde{Y}_d \phi {\cal D} + \mbox{\rm h.c.}~.\nnb
\eea
Here $M$ and $N$ running over 0,1,2,3,5 are the five--dimensional Lorentz
indices, $W_{MN}^a= {\partial}_M W_N^a - {\partial}_N W_M^a + \tilde{g}
\varepsilon^{abc} W_M^b W_N^c$ are the field strength tensor for the
$SU(2)_L$ electroweak gauge group, $B_{MN}={\partial}_M B_N -
{\partial}_N B_M$ are that of the $U(1)$ group, and all fields depend both on
$x$ and $y$. The covariant derivative is defined as ${\cal D}_M =
{\partial}_M - i \tilde{g} W_M^a T^a - i\tilde{g}^\prime B_M Y$, where 
$\tilde{g}$ and $\tilde{g}^\prime$ are the five--dimensional gauge couplings
for the $SU(2)_L$ and $U(1)$ groups. The five--dimensional $\Gamma_M$
matrices are defined as $\Gamma^\mu=\gamma^\mu~,~\mu=0,1,2,3$ and $\Gamma^5
= i \gamma^5$.

In the case o a single extra dimension with coordinate $x_5=y$ compactified
on a circle of radius $R$, a field $F(x,y)$ would be periodic function of
$y$, hence can be written as
\bea
F(x,y)=\sum_{n=-\infty}^{+\infty} F_n(x) e^{iny/R}~.\nnb
\eea

The Fourier expansion of the fields are
\bea
B_\mu(x,y) \es \frac{1}{\sqrt{2 \pi R}} B_\mu^{(0)} + 
\frac{1}{\sqrt{\pi R}} \sum_{n=1}^{\infty} B_\mu^{(0)} (x) 
\cos \left(\frac{ny}{R} \right)~, \nnb \\
B_5(x,y) \es \frac{1}{\sqrt{\pi R}} \sum_{n=1}^{\infty} B_5^{(n)} 
\sin \left(\frac{ny}{R} \right)~, \nnb \\
{\cal Q}(x,y) \es \frac{1}{\sqrt{2 \pi R}} {\cal Q}_L^{(0)} +
\frac{1}{\sqrt{\pi R}} \sum_{n=1}^{\infty} \left[
{\cal Q}_L^{(n)} \cos \left(\frac{ny}{R} \right) +
{\cal Q}_R^{(n)} \sin \left(\frac{ny}{R} \right) \right]~,\nnb \\
U({\cal  D})(x,y) \es \frac{1}{\sqrt{2 \pi R}} U_R^{(0)} +
\frac{1}{\sqrt{\pi R}} \sum_{n=1}^{\infty} \left[
U_R^{(n)} \cos \left(\frac{ny}{R} \right) + 
U_L^{(n)} \sin \left(\frac{ny}{R} \right) \right]~.\nnb
\eea
Under parity transformation $P_5:y \rar -y$ fields having a correspondent in
the four--dimensional SM should be even, so that their zero--mode in the
KK can be interpreted as the ordinary SM field. Fields having no
correspondent in the SM should be odd.    
From this expansion we see that fifth component of the vector field is odd
under $P_5$ transformation.

One important property of ACD model is the KK parity is conserved. The
parity conservation leads to the result that there is no tree level
contribution of KK modes in low energy processes (at the scale 
$\mu \ll 1/R$) and single KK excitation cannot be produced in ordinary
particle interaction. Finally note that in the ACD model there are three
additional physical scalar modes $a_n^{(0)}$ and $a_n^\pm$.

The zero--mode is either right--handed or left--handed. The nonzero--modes
come in chiral pair. This chirality is a consequence of the orbifold
boundary conditions.

Lagrangian of the ACD model can be obtained by integrating over $x_5=y$
\bea
{\cal L}_4 (x) = \int_0^{2\pi R} {\cal L}_5 (x,y) dy~.\nnb
\eea
Note that the zero--mode remains massless unless we apply the Higgs
mechanism. All fields in the four--dimensional Lagrangian receive the KK
mass $n/R$ on account of the derivative operator ${\partial}_5$ acting on
them.The relevant Feynman rules are derived in \cite{R7622} and for more
details about the ACD model we refer the interested reader to \cite{R7622}
and \cite{R7623}.

After this preliminary introduction, let us come back and discuss the main
problem. In this section we present the matrix element of $\Lambda_b \rar
\Lambda \ell^+ \ell^-$ decay, as well as expressions of the branching ratio,
forward--backward asymmetry and lepton polarizations.

At quark level, $\Lambda_b \rar \Lambda \ell^+ \ell^-$ decay is described by
$b \rar s \ell^+ \ell^-$ transition. Effective Hamiltonian governing this
transition in the SM with $\Delta B=-1$, $\Delta S = 1$ is described in
terms of a set of local operators
\bea
\label{e7601}
{\cal H} = \frac{4 G_F}{\sqrt{2}} V_{tb} V^\ast_{ts} \sum_1^{10} C_i(\mu)
{\cal O}_i(\mu)~,
\eea
where $G_F$ is the Fermi constant, $V_{ij}$ are the elements of the
Cabibbo--Kobayashi--Maskawa (CKM) matrix. Explicit forms of the operators,
which are written in terms of quark and gluon fields can be found in
\cite{R7624}.

The Wilson coefficients in (\ref{e7601}) have been computed at NNLO in the
SM in \cite{R7625}. At NLO the coefficients are calculated for the ACD model
including the effects of KK modes, in \cite{R7622} and \cite{R7623}, which
we have used in our calculations. It should be noted here that, there does not
appear any new operator in the ACD model, and therefore, new effects are
implemented by modifying the Wilson coefficients existing in the SM, if we
neglect the contributions of the scalar fields, which are indeed very small. 

At $\mu={\cal O}(m_W)$ level, only $C_2^{(0)}$, $C_7^{(0)(m_W)}$, $C_8^{(0)(m_W)}$,
$C_9^{(0)(m_W)}$ and $C_{10}^{(0)(m_W)}$ are different from zero, and the
remaining coefficients are all zero.

In the following we do not consider the contribution to $\Lambda_b \rar
\Lambda \ell^+ \ell^-$ decay from the lepton pair being created from
$\bar{c}c$ resonance due to the ${\cal O}_2$ operators. It can be removed by
applying appropriate cuts to invariant dilepton mass around mass of the
resonance. 

In the SM, at quark level, $\Lambda_b \rar \Lambda \ell^+ \ell^-$ decay is 
described with the help of the operators $C_7$, $C_9$ and $C_{10}$ as
follows:
\bea
\label{e7602}
{\cal H}_{eff} \es \frac{G_F}{4 \sqrt{2}} V_{tb} V^\ast_{ts} \Big\{
C_7 \bar{s} i \sigma_{\mu\nu} (1+ \gamma_5) q^\nu \bar{\ell}\gamma^\mu \ell
+ C_9 \bar{s} \gamma_\mu (1-\gamma_5) b \bar{\ell} \gamma^\mu \ell \nnb \\
\ar C_{10} \bar{s} \gamma_\mu (1-\gamma_5) b \bar{\ell} \gamma^\mu 
\gamma_5 \ell \Big\}~.
\eea   

The renormalization scheme independent coefficient 
$C_7^{(0)eff}$ \cite{R7626} is given by
\bea
\label{e7603} 
C_7^{(0)eff}(\mu_b) \es \eta^{\frac{16}{23}}
C_7^{(0)}(\mu_W)+ \frac{8}{3} \left( \eta^{\frac{14}{23}} 
-\eta^{\frac{16}{23}} \right) C_8^{(0)}(\mu_W)+C_2^{(0)}(\mu_W) \sum_{i=1}^8
h_i \eta^{\alpha_i}~, 
\eea
where 
\bea
\eta \es \frac{\alpha_s(\mu_W)} {\alpha_s(\mu_b)}~, \nnb
\eea
and 
\bea C_2^{(0)}(\mu_W)=1~,~~
C_7^{(0)}(\mu_W)=-\frac{1}{2} D^\prime(x_t,1/R)~,~~
C_8^{(0)}(\mu_W)=-\frac{1}{2} E^\prime(x_t,1/R)~, \nnb
\eea
with the superscript $(0)$ referring to leading log approximation, 
and coefficients $a_i$ and $h_i$ are given as
\bea 
a_1 \es \frac{14}{23}~,~~a_2=\frac{16}{23}~,~~a_3=\frac{6}{23}~,~~
a_4=-\frac{12}{23}~,~~a_5= 0.4086~,~~a_6=-0.4230~,\nnb \\
a_7 \es -0.8994~,a_8=0.1456 \nnb \\ \nnb \\
h_1 \es 2.2996~,~~h_2=-1.0880~,~~h_3=-\frac{3}{7}~,~~h_4=-\frac{1}{14}~,~~
h_5=-0.6494~,~~h_6=-0.0380~, \nnb \\
h_7 \es -0.0185~,~~h_8=-0.0057 \nnb
\eea

The functions $D^\prime$ and $E^\prime$, which describe electromagnetic and 
chromomagnetic penguins, respectively, are calculated in \cite{R7622} and
\cite{R7623}, and lead to the following results (see also \cite{R7624})

\bea
\label{e7604}
D^\prime_0(x_t) \es - \frac{(8 x_t^3+5 x_t^2-7 x_t)}{12 (1-x_t)^3}
+ \frac{x_t^2(2-3 x_t)}{2(1-x_t)^4}\ln x_t~, \\ \nnb \\
\label{e7605}
E^\prime_0(x_t) \es - \frac{x_t(x_t^2-5 x_t-2)}{4 (1-x_t)^3} +
\frac{3 x_t^2}{2 (1-x_t)^4}\ln x_t~, \\ \nnb \\
\label{e7606}
D^\prime_n(x_t,x_n) \es  \frac{x_t[-37+44 x_t+17 x_t^2+6 
x_n^2(10-9 x_t+3 x_t^2) -3 x_n (21-54 x_t+17 x_t^2)]}
{36 (x_t-1)^3 }\nnb \\
\ar \frac{x_n(2-7 x_n+3 x_n^2)}{6} \ln \frac{x_n}{1+ x_n} \nnb \\
\ar \frac{ (2-x_n-3 x_t)[x_t+3 x_t^2+x_n^2(3+x_t) - 
x_n + (10-x_t)x_n x_t] }
{6 (x_t-1)^4} \ln \frac{x_n+x_t}{1+x_n}~, \\ \nnb \\
\label{e7607}
E^\prime_n(x_t,x_n)\es \frac{x_t[-17-8 x_t+x_t^2-3 x_n(21-6 x_t+x_t^2)-6
x_n^2(10-9 x_t+3 x_t^2)]}{12 (x_t-1)^3} \nnb \\
\ek \frac{1}{2} x_n(1+x_n)(-1+3 x_n)\ln \frac{x_n}{1+ x_n} \nnb \\
\ar \frac{(1+x_n) [x_t+3 x_t^2+x_n^2(3+x_t)-x_n  +(10- x_t) x_n x_t ] }
{2(x_t-1)^4} \ln \frac{x_n+x_t}{1+x_n}~,
\eea
where the functions with and without $x_n$ correspond to the KK and SM
excitation contributions, respectively. Using the prescription presented in
\cite{R7622} and \cite{R7623}, and summing over $n$, we get the following 
expressions:

\bea
\label{e7608}
\sum_{n=1}^{\infty}D^\prime_n(x_t,x_n) \es  
\frac{x_t[37 - x_t(44+17 x_t)]}{72 (x_t-1)^3} \nnb \\
\ar \frac{\pi M_W R}{12} \Bigg[ \int_0^1 dy \, (2 y^{1/2}+7
y^{3/2}+3 y^{5/2}) \, \coth (\pi M_WR \sqrt{y}) \nnb \\
\ek \frac{x_t (2-3 x_t) (1+3 x_t)}{(x_t-1)^4}J(R,-1/2)\nnb \\
\ek \frac{1}{(x_t-1)^4} \{ x_t(1+3 x_t)+(2-3 x_t)
[1-(10-x_t)x_t] \} J(R, 1/2)\nnb \\
\ek \frac{1}{(x_t-1)^4} [ (2-3 x_t)(3+x_t) + 1 - (10-x_t) x_t] 
J(R, 3/2)\nnb \\
\ek \frac{(3+x_t)}{(x_t-1)^4} J(R,5/2)\Bigg]~, \\ \nnb \\
\label{e7609}
\sum_{n=1}^{\infty}E^\prime_n(x_t,x_n)\es \frac{x_t[17+(8-x_t)x_t]}
{24 (x_t-1)^3} \nnb \\
\ar \frac{\pi M_W R}{4} \Bigg[\int_0^1 dy \, (y^{1/2}+
2 y^{3/2}-3 y^{5/2}) \, \coth (\pi M_WR \sqrt{y}) \nnb \\
\ek {x_t(1+3 x_t) \over (x_t-1)^4}J(R,-1/2)\nnb \\
\ar \frac{1}{(x_t-1)^4} [ x_t(1+3 x_t) - 1 + (10-x_t)x_t] J(R, 1/2)\nnb \\
\ek \frac{1}{(x_t-1)^4} [(3+x_t)-1+(10-x_t)x_t) ]J(R, 3/2)\nnb \\
\ar{(3+x_t) \over  (x_t-1)^4} J(R,5/2)\Bigg]~,
\eea 
where 
\bea 
\label{e76010}
J(R,\alpha)=\int_0^1 dy \, y^\alpha \left[ \coth (\pi
M_W R \sqrt{y})-x_t^{1+\alpha} \coth(\pi m_t R \sqrt{y}) \right]~.
\eea

The Wilson coefficient $C_9$
in the ACD model and in the NDR scheme is 
\bea
\label{e7611}
C_9(\mu)=P_0^{NDR}+{Y(x_t,1/R) \over \sin^2 \theta_W} -4
Z(x_t,1/R)+P_E E(x_t,1/R)~, 
\eea 
where $P_0^{NDR}=2.60 \pm 0.25$
\cite{R7627} and the last term is numerically negligible.
The functions $Y(x_t,1/R)$ and $Z(x_t,1/R)$ are defined as:
\bea 
\label{e7612}
Y(x_t,1/R) \es Y_0(x_t)+\sum_{n=1}^\infty C_n(x_t,x_n)~, \nnb \\
Z(x_t,1/R) \es Z_0(x_t)+\sum_{n=1}^\infty C_n(x_t,x_n)~, 
\eea 
with
\bea 
\label{e7613}
Y_0(x_t) \es \frac{x_t}{8} \left[ \frac{x_t -4}{x_t -1}+
\frac{3 x_t}{(x_t-1)^2} \ln x_t \right]~,\nnb \\ \nnb \\
Z_0(x_t) \es \frac{18 x_t^4-163 x_t^3+259 x_t^2 -108
x_t}{144 (x_t-1)^3} \nnb \\
\ar \left[\frac{32 x_t^4-38 x_t^3-15 x_t^2+18 x_t}{72
(x_t-1)^4} - \frac{1}{9}\right] \ln x_t \\ \nnb \\
\label{e7614}
C_n(x_t,x_n) \es \frac{x_t}{8 (x_t-1)^2} \left[x_t^2-8 x_t+7+(3 +3
x_t+7 x_n-x_t x_n)\ln \frac{x_t+x_n}{1+x_n}\right]~, 
\eea 
and 
\bea
\label{e7615}
\sum_{n=1}^\infty C_n(x_t,x_n) = \frac{x_t(7-x_t)}{
16 (x_t-1)} - \frac{\pi M_W R x_t}{16 (x_t-1)^2}
\left[3(1+x_t)J(R,-1/2)+(x_t-7)J(R,1/2) \right]~.
\eea

Finally the Wilson coefficient $C_{10}$, which is scale independent, is given by:
\bea 
\label{e7616}
C_{10}= - \frac{Y(x_t,1/R)}{\sin^2 \theta_W}~.
\eea

With these coefficients and the operators in (\ref{e7601}) the inclusive 
$b \to s \ell^+ \ell^-$ transitions have been studied in 
\cite{R7622,R7623}.  

The amplitude of the exclusive $\Lambda_b \rar \Lambda\ell^+ \ell^-$ decay
is obtained by calculating the matrix element of the effective Hamiltonian for the 
$b \rar s \ell^+ \ell^-$ transition between initial and final
baryon states $\lla \Lambda \vel {\cal H}_{eff} \ver \Lambda_b \rra$.
It follows from Eq. (\ref{e7616}) that the matrix elements
\bea
\label{e7117}
&&\lla \Lambda \vel \bar s \gamma_\mu (1 - \gamma_5) b \ver \Lambda_b
\rra~,\nnb \\
&&\lla \Lambda \vel \bar s \sigma_{\mu\nu} (1 + \gamma_5) b \ver \Lambda_b
\rra~,
\eea
are needed in order to calculate
the $\Lambda_b \rar \Lambda\ell^+ \ell^-$ decay amplitude.

These matrix elements parametrized in terms of the form factors are 
as follows (see \cite{R7614,R7628})
\bea
\label{e7618}
\lla \Lambda \vel \bar s \gamma_\mu b \ver \Lambda_b \rra  
\es \bar u_\Lambda \Big[ f_1 \gamma_\mu + i f_2 \sigma_{\mu\nu} q^\nu + f_3  
q_\mu \Big] u_{\Lambda_b}~,\\
\label{e7619}
\lla \Lambda \vel \bar s \gamma_\mu \gamma_5 b \ver \Lambda_b \rra
\es \bar u_\Lambda \Big[ g_1 \gamma_\mu \gamma_5 + i g_2 \sigma_{\mu\nu}
\gamma_5 q^\nu + g_3 q_\mu \gamma_5\Big] u_{\Lambda_b}~,
\eea
where $q= p_{\Lambda_b} - p_\Lambda$. 

The form factors of the magnetic dipole operators are defined as 
\bea
\label{e7620}
\lla \Lambda \vel \bar s i \sigma_{\mu\nu} q^\nu  b \ver \Lambda_b \rra
\es \bar u_\Lambda \Big[ f_1^T \gamma_\mu + i f_2^T \sigma_{\mu\nu} q^\nu
+ f_3^T q_\mu \Big] u_{\Lambda_b}~,\nnb \\
\lla \Lambda \vel \bar s i \sigma_{\mu\nu}\gamma_5  q^\nu  b \ver \Lambda_b \rra
\es \bar u_\Lambda \Big[ g_1^T \gamma_\mu \gamma_5 + i g_2^T \sigma_{\mu\nu}
\gamma_5 q^\nu + g_3^T q_\mu \gamma_5\Big] u_{\Lambda_b}~.
\eea

Using the identity 
\bea
\sigma_{\mu\nu}\gamma_5 = - \frac{i}{2} \epsilon_{\mu\nu\alpha\beta}
\sigma^{\alpha\beta}~,\nnb
\eea
the following relations between the form factors are obtained:
\bea
\label{e7621}
f_1^T \es - \frac{q^2}{m_{\Lambda_b} - m_\Lambda} f_3^T~,\nnb \\
g_1^T \es \frac{q^2}{m_{\Lambda_b} + m_\Lambda} g_3^T~.
\eea 

Using these definitions of the form factors, for the matrix element
of the $\Lambda_b \rar \Lambda\ell^+ \ell^-$ we get
\bea
\label{e7622}
{\cal M} \es \frac{G \alpha}{4 \sqrt{2}\pi} V_{tb}V_{ts}^\ast \frac{1}{2} \Bigg\{
\bar{\ell} \gamma_\mu (1-\gamma_5) \ell \, 
\bar{u}_\Lambda \Big[ (A_1 - D_1) \gamma_\mu (1+\gamma_5) +
(B_1 - E_1) \gamma_\mu (1-\gamma_5) \nnb \\
\ar i \sigma_{\mu\nu} q^\nu \Big( (A_2 - D_2) (1+\gamma_5) +
(B_2 - E_2) (1-\gamma_5) \Big) \nnb \\
\ar q_\mu \Big( (A_3 - D_3) (1+\gamma_5) + (B_3 - E_3) (1-\gamma_5)
\Big) \Big] u_{\Lambda_b} \nnb \\
\ar \bar{\ell} \gamma_\mu (1+\gamma_5) \ell \, 
\bar{u}_\Lambda \Big[ (A_1 + D_1) \gamma_\mu (1+\gamma_5) +
(B_1 + E_1) \gamma_\mu (1-\gamma_5) \nnb \\
\ar i \sigma_{\mu\nu} q^\nu \Big( (A_2 + D_2) (1+\gamma_5) +
(B_2 + E_2) (1-\gamma_5) \Big) \nnb \\
\ar q_\mu \Big( (A_3 + D_3) (1+\gamma_5) + (B_3 + E_3) (1-\gamma_5) \Big)
\Big] u_{\Lambda_b} \Bigg\}~,
\eea
where
\bea
\label{e7623}
A_1 \es \frac{1}{q^2}\ga f_1^T-g_1^T \dr (-2 m_s C_7) + \frac{1}{q^2}\ga
f_1^T+g_1^T \dr (-2 m_b C_7) + \ga f_1-g_1 \dr C_9^{eff}~,\nnb \\
A_2 \es A_1 \ga 1 \rar 2 \dr ~,\nnb \\
A_3 \es A_1 \ga 1 \rar 3 \dr ~,\nnb \\
B_1 \es A_1 \ga g_1 \rar - g_1;~g_1^T \rar - g_1^T \dr ~,\nnb \\
B_2 \es B_1 \ga 1 \rar 2 \dr ~,\nnb \\
B_3 \es B_1 \ga 1 \rar 3 \dr ~,\nnb \\
D_1 \es C_{10} \ga f_1-g_1 \dr~,\nnb \\
D_2 \es D_1 \ga 1 \rar 2 \dr ~, \\
D_3 \es D_1 \ga 1 \rar 3 \dr ~,\nnb \\
E_1 \es D_1 \ga g_1 \rar - g_1 \dr ~,\nnb \\
E_2 \es E_1 \ga 1 \rar 2 \dr ~,\nnb \\
E_3 \es E_1 \ga 1 \rar 3 \dr ~.\nnb
\eea

From these expressions it follows
that $\Lambda_b \rar\Lambda \ell^+\ell^-$ decay is described in terms of  
many form factors. It is shown in \cite{R7629} that Heavy Quark Effective
Theory reduces
the number of independent form factors to two ($F_1$ and
$F_2$) irrelevant of the Dirac structure
of the corresponding operators, i.e., 
\bea
\label{e7624}
\lla \Lambda(p_\Lambda) \vel \bar s \Gamma b \ver \Lambda(p_{\Lambda_b})
\rra = \bar u_\Lambda \Big[F_1(q^2) + \not\!v F_2(q^2)\Big] \Gamma
u_{\Lambda_b}~,
\eea
where $\Gamma$ is an arbitrary Dirac structure and
$v^\mu=p_{\Lambda_b}^\mu/m_{\Lambda_b}$ is the four--velocity of
$\Lambda_b$. Comparing the general form of the form factors given in Eqs.
(\ref{e7617})--(\ref{e7620}) with (\ref{e7623}), one can
easily obtain the following relations among them (see also
\cite{R7614,R7615,R7628})
\bea
\label{e7625}
g_1 \es f_1 = f_2^T= g_2^T = F_1 + \sqrt{\hat{r}_\Lambda} F_2~, \nnb \\
g_2 \es f_2 = g_3 = f_3 = \frac{F_2}{m_{\Lambda_b}}~,\nnb \\
g_1^T \es f_1^T = \frac{F_2}{m_{\Lambda_b}} q^2~,\nnb \\
g_3^T \es \frac{F_2}{m_{\Lambda_b}} \ga m_{\Lambda_b} + m_\Lambda \dr~,\nnb \\
f_3^T \es - \frac{F_2}{m_{\Lambda_b}} \ga m_{\Lambda_b} - m_\Lambda \dr~,
\eea
where $\hat{r}_\Lambda=m_\Lambda^2/m_{\Lambda_b}^2$.

In what follows, we will be looking for the possible manifestation of UED
theory in branching ratio, as well as in lepton polarizations. For this
purpose, we present the decay rate for the $\Lambda_b \rar \Lambda \ell^+
\ell^-$ taking into account lepton polarizations. 

In th rest frame of lepton $\ell^-$ the unit vectors along the longitudinal,
normal and transversal components of the $\ell^-$ are chosen as:
\bea
\label{e7626}
s^{-\mu}_L \es \ga 0,\vec{e}_L^{\,-}\dr =
\ga 0,\frac{\vec{p}_-}{\vel\vec{p}_- \ver}\dr~, \nnb \\
s^{-\mu}_N \es \ga 0,\vec{e}_N^{\,-}\dr = \ga 0,\frac{\vec{p}_\Lambda\times
\vec{p}_-}{\vel \vec{p}_\Lambda\times \vec{p}_- \ver}\dr~, \nnb \\
s^{-\mu}_T \es \ga 0,\vec{e}_T^{\,-}\dr = \ga 0,\vec{e}_N^{\,-}
\times \vec{e}_L^{\,-} \dr~,
\eea
where $\vec{p}_-$ and $\vec{p}_\Lambda$ are the three--momenta of the
leptons $\ell^-$ and $\Lambda$ baryon in the
center of mass frame (CM) of $\ell^- \,\ell^+$ system, respectively.

The longitudinal component of the lepton polarization is boosted to the CM
frame of the lepton pair by Lorentz transformation, yielding
\bea
\label{e7627}
\ga s^{-\mu}_L \dr_{CM} \es \ga \frac{\vel\vec{p}_- \ver}{m_\ell}~,
\frac{E_\ell \,\vec{p}_-}{m_\ell \vel\vec{p}_- \ver}\dr~,
\eea
where $E_\ell$ and $m_\ell$ are the energy and mass of $\ell^-$ in the 
CM frame. The remaining two unit vectors $s_N^{-\mu}$, $s_T^{-\mu}$ 
are unchanged under Lorentz boost.

The differential decay rate for $\Lambda_b \rar \Lambda \ell^+ \ell^-$ decay
along any spin direction $\vec{\xi}^{\;\pm}$ along the $\ell^\pm$ can be written
as:
\bea
\label{e7628}
\frac{d\Gamma(\vec{\xi}^{\;\mp})}{d\hat{s}} = \frac{1}{2}
\ga \frac{d\Gamma}{d\hat{s}}\dr_0
\Bigg[ 1 + \Bigg( P_L^\mp \vec{e}_L^{\;\mp} + P_N^\mp
\vec{e}_N^{\;\mp} + P_T^\mp \vec{e}_T^{\;\mp} \Bigg) \cdot
\vec{\xi}^{\;\mp} \Bigg]~,
\eea
where $\ga d\Gamma/d\hat{s} \dr_0$ corresponds to the unpolarized differential
decay rate, $\hat{s}=q^2/m_{\Lambda_b}^2$ and    
$P_L$, $P_N$ and $P_T$ represent the longitudinal, normal and 
transversal polarizations of $\ell$, respectively, and has the following
form:
\bea
\label{e7629}
\ga \frac{d \Gamma}{d\hat{s}}\dr_0 = \frac{G^2 \alpha^2}{8192 \pi^5}
\vel V_{tb} V_{ts}^\ast \ver^2 \lambda^{1/2}(1,r,\hat{s}) v 
\Big[{\cal T}_0(\hat{s}) +\frac{1}{3} {\cal T}_2(\hat{s}) \Big]~, 
\eea
where 
$\lambda(1,r,\hat{s}) = 1 + r^2 + \hat{s}^2 - 2 r - 2 \hat{s} - 2 r\hat{s}$
is the triangle function and $v=\sqrt{1-4m_\ell^2/q^2}$ is the lepton
velocity.

The polarizations $P_L$, $P_N$ and $P_T$ are defined as:
\bea
\label{e7629}
P_i^{(\mp)}(\hat{s}) = \frac{\ds{\frac{d \Gamma}{d\hat{s}}
                   (\vec{\xi}^{\;\mp}=\vec{e}_i^{\;\mp}) -
                   \frac{d \Gamma}{d\hat{s}}
                   (\vec{\xi}^{\;\mp}=-\vec{e}_i^{\;\mp})}}
              {\ds{\frac{d \Gamma}{d\hat{s}}
                   (\vec{\xi}^{\;\mp}=\vec{e}_i^{\;\mp}) +
                  \frac{d \Gamma}{d\hat{s}}
                  (\vec{\xi}^{\;\mp}=-\vec{e}_i^{\;\mp})}}~, \nnb
\eea
where $i = L,N,T$.

One of the efficient tools for establishing new physics effects is study of
the forward--backward asymmetry ${\cal A}_{FB}$ which is defined as
\bea
{\cal A}_{FB} = \frac{\ds{\int_0^1\frac{d\Gamma}{d\hat{s}dz}}\,dz -
\ds{\int_{-1}^0\frac{d\Gamma}{d\hat{s}dz}}\,dz}
{\ds{\int_0^1\frac{d\Gamma}{d\hat{s}dz}}\,dz +
\ds{\int_{-1}^0\frac{d\Gamma}{d\hat{s}dz}}\,dz}~,\nnb
\eea
where $z=\cos\theta$ dependence of the differential decay rate can be
implemented by making the replacement
\bea
{\cal T}_0(\hat{s}) + \frac{1}{3}{\cal T}_2(\hat{s}) \rar 
{\cal T}_0(\hat{s})+{\cal T}_1(\hat{s}) z +{\cal T}_2(\hat{s}) z^2~,\nnb
\eea
on the right-hand side of Eq. (\ref{e7628}), and $\theta$ is the angle
between $\Lambda_b$ and
$\ell^-$ in the CM of leptons. Explicit expressions of ${\cal T}_0(\hat{s})$,
${\cal T}_1(\hat{s})$ and ${\cal T}_2(\hat{s})$ can be found in \cite{R7614}.

It is well known that in the $B \rar K^\ast \ell^+ \ell^-$ decay the 
zero--position of ${\cal A}_{FB}$ is practically independent of the 
form factors \cite{R7628}. For this reason, determination of 
zero--position of ${\cal A}_{FB}$, as well as it magnitude,is very 
promising in looking for new physics beyond the SM.
Note also that the combined analysis of the lepton polarizations can give
additional information about the existence of new physics since in the SM
$P_L^+ + P_L^- = 0$, $P_N^+ + P_N^- = 0$ and $P_T^+ - P_T^- \approx 0$ (in
the $m_\ell \rar 0$ limit). Therefore any nonzero value resulting from these
combined polarizations can be considered as an confirmation of new physics.

\section{Numerical analysis}

In this section we present our numerical results for the 
polarization asymmetries $P_L$, $P_N$ and $P_T$ when one of the leptons is
polarized. The values of the input parameters we use in our
calculations are: $\vel V_{tb} V_{ts}^\ast \ver = 0.0385$, 
$m_\tau = 1.77~GeV$, $m_\mu = 0.106~GeV$.
$m_b = 4.8~GeV$ \cite{R7630}, $m_t=172.7~GeV$ \cite{R7631} and $\tau_{B_0} =
1.527 \pm 0.008$.

From the expressions of asymmetries it follows that the form
factors are the main and the most important input parameters necessary in
the numerical calculations. The calculation of the form factors of $\Lambda_b
\rar \Lambda$ transition does not exist at present.
But, we can use the results from QCD sum rules
in corporation with HQET \cite{R7629,R7632}. We noted earlier that,
HQET allows us to establish relations among the form factors and reduces
the number of independent form factors into two. 
In \cite{R7629,R7632}, the $q^2$ dependence of these form factors
are given as follows
\bea
F(\hat{s}) = \frac{F(0)}{\ds 1-a_F \hat{s} + b_F \hat{s}^2}~. \nnb
\eea
The values of the parameters $F(0),~a_F$ and $b_F$ are given in table 1.
\begin{table}[h]    
\renewcommand{\arraystretch}{1.5}
\addtolength{\arraycolsep}{3pt}
$$
\begin{array}{|l|ccc|}  
\hline
& F(0) & a_F & b_F \\ \hline
F_1 &
\phantom{-}0.462 & -0.0182 & -0.000176 \\
F_2 &
-0.077 & -0.0685 &\phantom{-}0.00146 \\ \hline
\end{array}
$$
\caption{Form factors for $\Lambda_b \rar \Lambda \ell^+ \ell^-$
decay in a three parameter fit.}
\renewcommand{\arraystretch}{1}
\addtolength{\arraycolsep}{-3pt}
\end{table}  

Note that the first analysis of the HQET structure of the $\Lambda_Q \rar
\Lambda_q$ transition is performed in \cite{R7633} (see also \cite{R7634}).

In order to have an idea about the sensitivity of our results to the
specific parametrization of the two form factors predicted by the QCD sum
rules in corporation with the HQET, we also have used another
parametrization of the form factors based on the pole model and compared the
results of both models. The second set of form factors which have the dipole
form predicted by the pole model are given as:
\bea
F_{1,2}(E_\Lambda) = N_{1,2} \left( \frac{\Lambda_{QCD}}{\Lambda_{QCD}+
E_\Lambda} \right)^2~,\nnb
\eea
where
\bea
E_\Lambda = \frac{m_{\Lambda_b}^2 - m_\Lambda^2 - q^2}
{2 m_{\Lambda_b}}~,\nnb
\eea
and $\Lambda_{QCD} = 0.2$, $N_1 = 52.32$ and $N_2 \simeq
-0.25 N_1$ \cite{R7635}.

In Figs. 1 and 2 we present the dependence of the branching ratio for
the $\Lambda_b \rar\Lambda \mu^+\mu^-$ and $\Lambda_b \rar\Lambda
\tau^+\tau^-$ decays on the compactification
parameter $1/R$, respectively. For a comparison, in these figures we also 
present the SM prediction for this decay using both set of form factors. 
From these figures we se that the
branching ratios are larger compared to the SM result for both
decay modes and practically they are insensitive to the form factors. 
The value of the branching ratio for $1/R=200~GeV$ 
is approximately $50\%$ larger compared to that of the SM prediction. But
this difference decreases as $1/R$ gets larger. Therefore measurement of the
value of branching ratio of the $\Lambda_b \rar\Lambda \ell^+\ell^-$, $\ell
= \mu,~\tau$ decays can serve as a good test for physics beyond the SM.

Analysis of the lepton polarizations leads to the following results:
\begin{itemize}

\item 
Maximum value of the difference between the SM and ADC models (for the
minimum value of $1/R=200~GeV$) predictions, as far as longitudinal
polarization is concerned, is about $10\%$.

\item   
Practically, there is no difference between the predictions of the SM and 
ACD models fo the $\tau$ lepton with longitudinal polarization.

\item
These two models lead to the same result for the $\mu$ lepton with
transversal polarization case.

\item
Up to $s=0.6$, the maximum difference in the predictions of these two models
is about $12\%$ for the $\tau$ lepton with transversal polarization case.  

\item
Normal polarization of lepton contributes very small and the difference
between predictions of the two models for this polarization can never be
measured in the experiments. For the $\tau$ lepton case  the difference
$(P_N)_{ACD} - (P_N)_{SM} \approx 0.5 \%$ is quite a challenging to be
measured.

\end{itemize}

From this discussion we conclude that measurement of polarizations of lepton
is not useful for establishing the UED models.
 
We can now discuss the prediction of the ACD model for the forward--backward
asymmetry. In Fig. 3 we show the dependence of of ${\cal A}_{FB}$ on
$\hat{s}$ at four fixed values of $1/R$ and SM, 
for the $\Lambda_b \rar\Lambda \mu^+\mu^-$ decay when the first set of form 
factors are considered.
For the sake of completeness, we also present in this figure the
forward--backward asymmetry prediction of SM. From this figure we see that
the zero--position of ${\cal A}_{FB}$ is sensitive to the compactification
parameter $1/R$, similar to the $B \rar K^\ast \ell^+ \ell^-$ decay case and 
in all cases it is shifted to the left compared to that of the SM prediction.
Therefore, experimental determination of zero--position ${\cal A}_{FB}$ can
give invaluable information about new physics effects. It should be noted
that the zero position of ${\cal A}_{FB}$ is practically insensitive to the
choice of form factors and it coincides for both sets.  

In conclusion, we have studied the rare $\Lambda_b \rar\Lambda \ell^+\ell^-$
decay in the ACD model with a single universal extra dimension. We
investigate the sensitivity of the branching ratio, lepton polarizations and
lepton forward--backward asymmetry in the $\Lambda_b \rar\Lambda \ell^+\ell^-$ 
decay to the compactification parameter $1/R$. We found that the branching
ratio and zero--position of the forward--backward asymmetry are very
sensitive to the presence of the compactification parameter $1/R$ and can be
useful for establishing new physics effects due to the compactification of
the fifth dimension. 
     
\section*{Acknowledgments}

One of the authors (T. M. A) is grateful to T\"{U}B\.{I}TAK for partially
support of this work under the project 105T131.

\newpage

\section*{Figure captions}
{\bf Fig. 1} The dependence of the branching ratio for the 
$\Lambda_b \rar \Lambda \mu^+ \mu^-$ decay on the compactification
parameter $1/R$, for the first set of form factors I and the second set of
form factors II, respctively. \\ \\
{\bf Fig. 2} The same as Fig. 1, but for the 
$\Lambda_b \rar \Lambda \tau^+ \tau^-$ decay. \\ \\
{\bf Fig. 3} The dependence of the forward--backward asymmetry ${\cal
A}_{FB}$ on $\hat{s}$ at four fixed values of $1/R$ and in the SM, for the
$\Lambda_b \rar \Lambda \mu^+ \mu^-$ decay.

\newpage

\begin{figure}
\vskip 3. cm
    \includegraphics{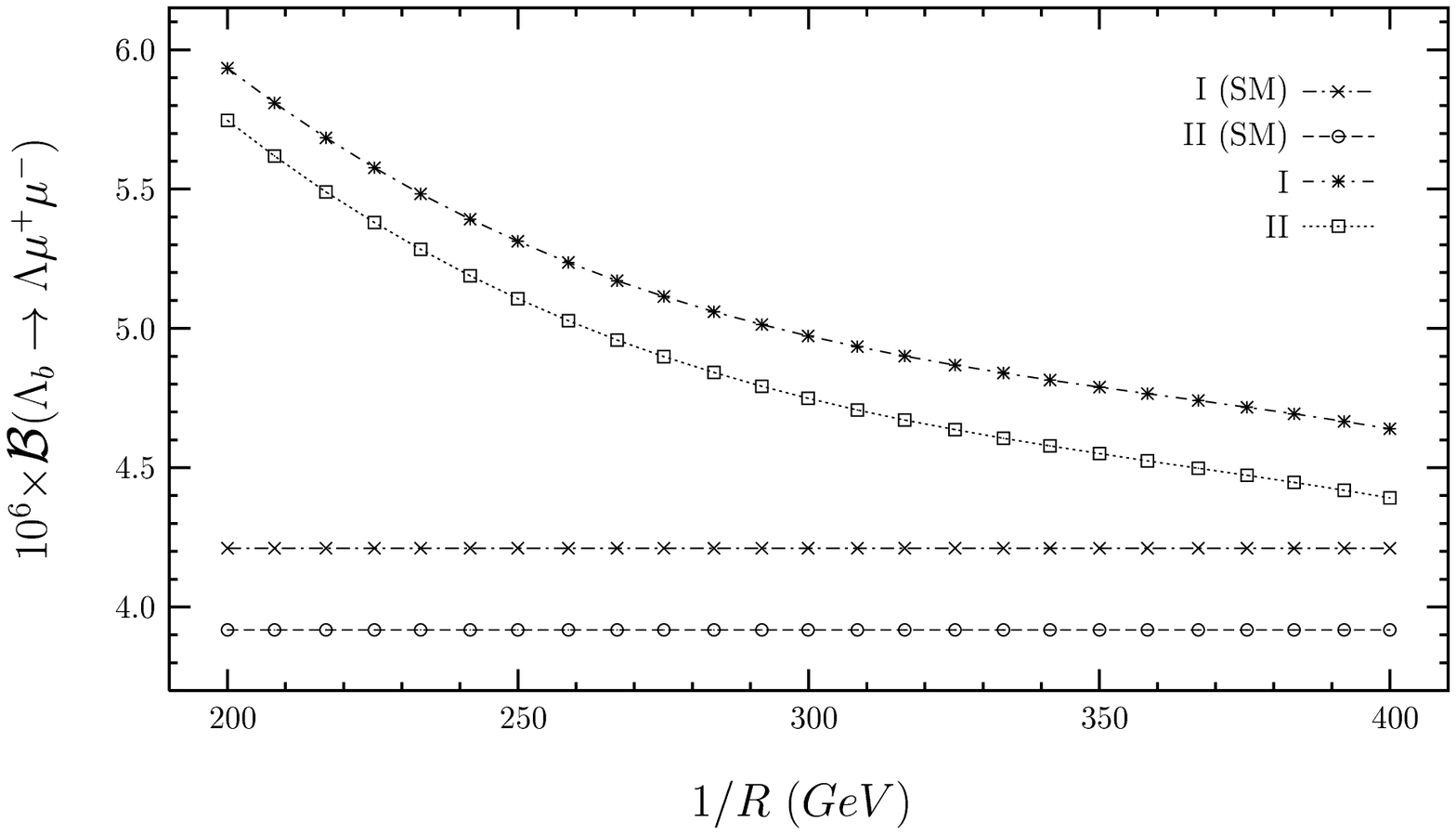}
\vskip 6.3cm
\caption{}
\end{figure}

\begin{figure}
\vskip 4.0 cm
    \includegraphics{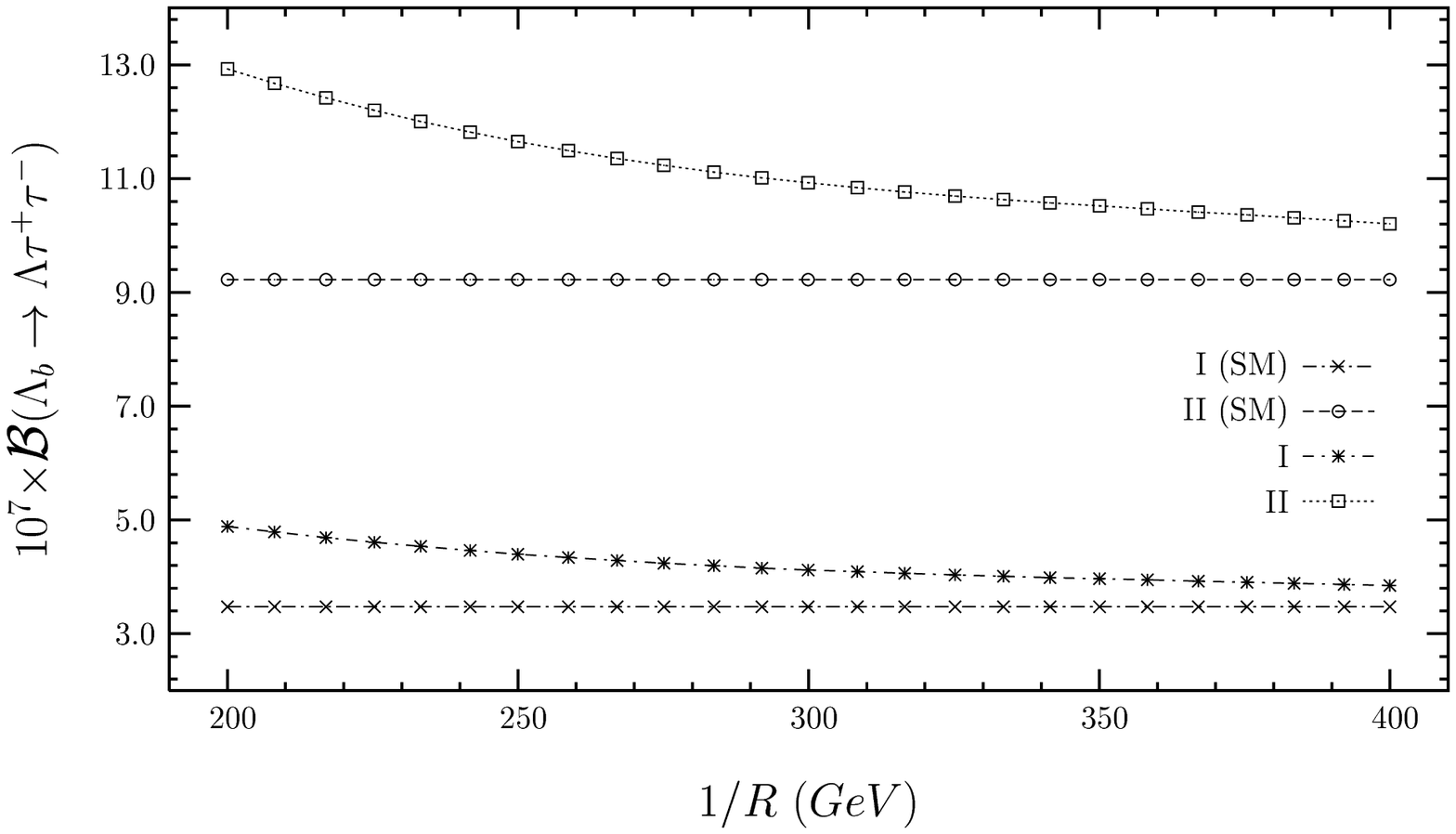}
\vskip 6.3 cm
\caption{}
\end{figure}

\begin{figure}
\vskip 3. cm
    \includegraphics{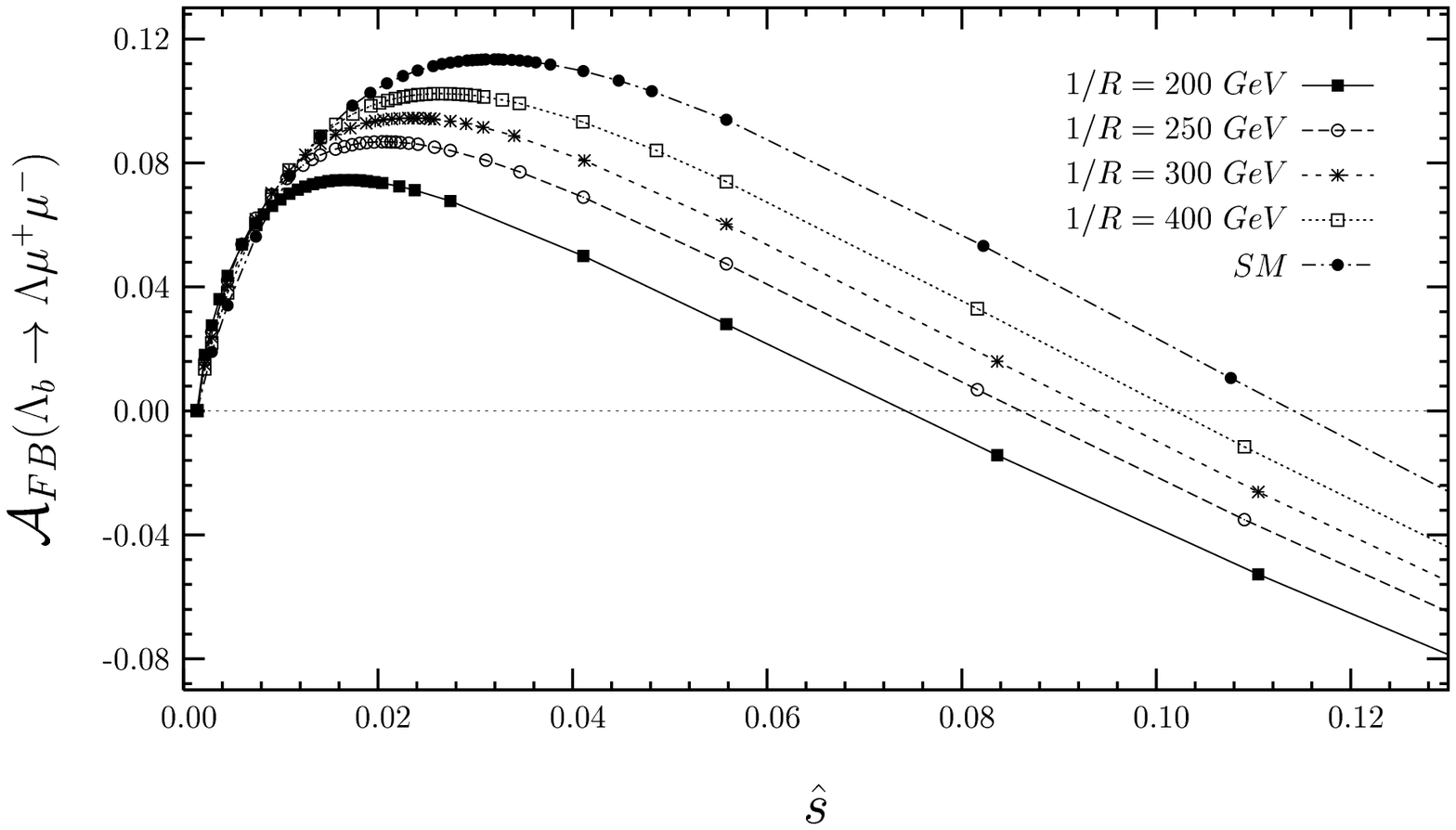}
\vskip 6.3cm
\caption{}
\end{figure}

\end{document}